\documentclass[fleqn,twoside,11pt]{article}

\usepackage{amssymb}
\usepackage{graphicx}
\usepackage{epsbox}
\usepackage{amsmath}

\begin{document}
\noindent
{\sf University of Shizuoka}

\hspace*{13cm} {\large US-05-05R}

\vspace{3mm}

\begin{center}

{\Large\bf  Permutation Symmetry S$_3$ and }\\[.1in]
{\Large\bf VEV Structure of Flavor-Triplet Higgs Scalars}

\vspace{2mm}
{\bf Yoshio Koide}

{\it Department of Physics, University of Shizuoka, 
52-1 Yada, Shizuoka 422-8526, Japan\\
E-mail address: koide@u-shizuoka-ken.ac.jp}

\date{\today}
\end{center}

\begin{abstract}
A model with flavor-triplet Higgs scalars $\phi_i$ ($i=1,2,3$)
is investigated under a permutation symmetry 
S$_3$  and its symmetry breaking. 
A possible S$_3$ breaking form of the Higgs potential whose
vacuum expectation values $v_i=\langle \phi_i\rangle$
satisfy a relation $v_1^2 +v_2^2 +v_3^2 =\frac{2}{3}
(v_1 +v_2 +v_3)^2$ is investigated, because if we suppose
a seesaw-like mass matrix model $M_e = m M^{-1} m$
with $m_{ij} \propto \delta_{ij} v_i$ and
$M_{ij} \propto \delta_{ij}$, such a model can lead
to the well-known charged lepton mass relation
$m_e +m_\mu +m_\tau = \frac{2}{3} (\sqrt{m_e}+\sqrt{m_\mu}
+\sqrt{m_\tau})^2$.
\end{abstract}

\vspace{3mm}

{\large\bf 1 \ Introduction}

Of the observed mass spectra of the fundamental particles, 
quarks and leptons, the charged lepton mass spectrum seems 
to give a promising clue to the unified understanding of 
quarks and leptons, because the observed charged lepton masses
satisfy a very simple mass 
relation \cite{Koide82,Koide83,Koide90} 
$$
m_e+m_{\mu}+m_{\tau}=
\frac{2}{3}\left( \sqrt{m_e}+\sqrt{m_\mu}+\sqrt{m_{\tau}} 
\right)^2 , 
\eqno(1.1)
$$
with remarkable precision.
The mass formula (1.1) can give an
excellent prediction of the tau lepton mass value 
$$
m_{\tau}=1776.97 \ {\rm MeV},
\eqno(1.2)
$$
from the observed electron and muon mass values 
\cite{PDG04}, $m_e=0.51099892$ MeV and 
$m_{\mu}=105.658369$ MeV ({\it c.f.} 
the observed value \cite{PDG04} 
$m_{\tau}=1776.99^{+0.29}_{-0.26}$ MeV).  
This excellent agreement seems to be beyond a matter
of accidental coincidence,
so that we should consider the origin of the mass 
formula (1.1) seriously.
Several authors \cite{Foot,Esposito,Li-Ma} have challenged to give 
an explanation of the mass formula (1.1) from a geometrical point
of view.
However, up to the present, the theoretical basis of 
the mass formula (1.1) is still not clear. 
(For a review, for example, see Ref.~\cite{Rivero, KoideLP05}.)

The charged lepton mass formula (1.1) has 
the following peculiar features:

\noindent(a) The mass formula is described in terms of the 
root squared masses $\sqrt{m_{ei}}$.

\noindent(b) The formula is well satisfied at a low energy scale 
rather than at a high energy scale. 

\noindent(c) The mass formula is invariant under the exchanges 
$\sqrt{m_{ei}} \leftrightarrow \sqrt{m_{ej}}$.

The feature (a) suggests that the charged lepton 
mass spectrum is given by a bilinear form 
on the basis of some mass-generation mechanism. 
For example, 
in Refs.~\cite{Koide90, KF96, KT96, Koide99}, the formula
(1.1) has been discussed on the basis of
a seesaw-like mechanism \cite{UnivSeesaw}: 
$$
M_e=m M^{-1}_E m^{\dagger} ,
\eqno(1.3)
$$
where $M_E$ is a heavy charged lepton mass matrix
$M_E \propto {\rm diag}(1,1,1)$, and $m$ is given by 
$m \propto {\rm diag}(v_1, v_2, v_3)$ ($v_i$ are
vacuum expectation values (VEVs) of flavor-triplet 
scalars $\phi_i$). 
This idea that mass spectrum 
is due not to the structure of the Yukawa coupling constants,
but to the VEV structure of Higgs scalars $\phi_i$ at a low 
energy scale is very attractive as an explanation of the 
feature (b).
We have to seek for a model where the VEVs $v_i$ satisfy 
the following relation
$$
v_1^2 + v_2^2 + v_3^2 = \frac{2}{3}
 \left( v_1 + v_2 + v_3 \right) ^2.
\eqno(1.4)
$$

The feature (c) suggests that the Higgs potential is 
invariant under a permutation symmetry S$_3$.
Such attempts to understand the charged lepton mass spectrum
from the VEV structure of flavor-triplet scalars are
found in Refs.\cite{Koide90,KT96,Koide99}.
The basic idea is as follows:
We consider the following S$_3$ invariant 
Higgs potential 
$$
V=\mu^2 \sum_{i}(\overline{\phi}_i \phi_i)
+\frac{1}{2} \lambda_1 \left[\sum_i(\overline{\phi}_i \phi_i)\right]^2
+ \lambda_2 (\overline{\phi}_{\sigma} \phi_{\sigma})
(\overline{\phi}_{\pi}\phi_{\pi} + \overline{\phi}_{\eta}\phi_{\eta}), 
\eqno(1.5)
$$
where 
$(\overline{\phi}_i \phi_i)=\phi^-_i \phi^+_i 
+\overline{\phi} \ ^0_i \phi^0_i$ ($i=1,2,3$), and 
$(\phi_\pi, \phi_\eta)$ and $\phi_\sigma$ are a doublet and
a singlet in the real basis of S$_3$, respectively:
$$
\begin{array}{l}
\phi_{\pi} = \frac{1}{\sqrt{2}}(\phi_1 - \phi_2)  , \\ 
\phi_{\eta} = \frac{1}{\sqrt{6}}(\phi_1 + \phi_2 -2\phi_3)  , \\
\phi_{\sigma} = \frac{1}{\sqrt{3}}(\phi_1 + \phi_2 +\phi_3)  .
\end{array}
\eqno(1.6)
$$
Although, in Refs.\cite{Koide90,KT96,Koide99}, we have regarded 
the fields $\phi_i$ as SU(2)$_L$ doublet scalars, we can also consider
another model which gives the seesaw form (1.3).
For example, by considering a model shown in Fig.~1, we can regard
the fields $\phi_i$ as SU(2)$_L$ singlet scalars.
Hereafter, the SU(2)$_L$ structure in the Higgs potential will be neglected.
The fields $\phi_i$ denote both cases, a case of the SU(2)$_L$ doublets and
a case of the SU(2)$_L$ singlets.

The conditions  that 
the potential (1.5) takes the minimum lead to the relation
for the VEVs $v_i\equiv \langle \phi^0_i \rangle$
$$
|v_{\sigma}|^2= |v_{\pi}|^2 + |v_{\eta}|^2 
= \frac{-\mu^2}{2\lambda_1 + \lambda_2}  .
\eqno(1.7)
$$
Therefore, from the relation 
$$
\overline{\phi}_1 \phi_1+ \overline{\phi}_2 \phi_2
+ \overline{\phi}_3 \phi_3 =
\overline{\phi}_\pi \phi_{\pi}+ \overline{\phi}_\eta \phi_{\eta}
+ \overline{\phi}_\sigma \phi_{\sigma}  ,
\eqno(1.8)
$$
we obtain
$$
|v_1|^2 + |v_2|^2 + |v_3|^2 
=|v_{\pi}|^2 + |v_{\eta}|^2 + |v_{\sigma}|^2 
= 2|v_{\sigma}|^2
= 2 \left( \frac{v_1 + v_2 + v_3}{\sqrt{3}}\right) ^2.
\eqno(1.9)
$$
Thus, we obtain the relation (1.4).

However, note that (i) the Higgs potential (1.5) which is 
invariant under the permutation symmetry S$_3$ is not
a general form of the S$_3$ invariant Higgs potential,
and (ii) it cannot give a relation between $v_\pi =\langle \phi_\pi
\rangle$ and $v_\eta =\langle \phi_\eta\rangle$, because
the S$_3$ invariant Higgs potential (1.5) is also invariant under
the permutation between $\phi_\pi$ and $\phi_\eta$
(we cannot determine the values $v_\pi$ and $v_\eta$
individually).
We have to investigate a potential term which violates a $\phi_\pi
\leftrightarrow \phi_\eta$ symmetry (also breaks the S$_3$ symmetry) 
but keeping the relation (1.4).

In order to obtain the charged lepton mass relation (1.1),
we have to build a model with a seesaw-type mass matrix
(1.3).
A recent attempt to build such a model will be found in 
Ref.\cite{Koide05}.
However, the purpose of the present paper is not to investigate
such a seesaw mass matrix model.
The purpose of the present paper is to discuss the Higgs 
potential form which gives the structure (1.4).

\vspace{5mm}
{\large\bf 2 \ S$_3$ symmetric Higgs potential}

In this section, we discuss a general form of the S$_3$
symmetric Higgs potential.
However, for mass terms, we confine ourselves to the case 
$\mu^2 \sum \bar{\phi}_i \phi_i$ as given in Eq.~(1.5).
We consider an S$_3$-invariant general form only for
the dimension-four terms.
In general, we have two scalars $(\bar{\phi}_\sigma \phi_\sigma)$
and $(\bar{\phi}_\pi \phi_\pi+\bar{\phi}_\eta \phi_\eta)$ and
one pseudo-scalar 
$(\bar{\phi}_\pi \phi_\eta -\bar{\eta}_\eta \phi_\eta)$
\cite{Haba-Yoshioka}, 
so that we write the general form of the S$_3$-invariant 
Higgs potential as follows:
$$
V= \mu^2 \left( \bar{\phi}_1 \phi_1+\bar{\phi}_2 \phi_2
+\bar{\phi}_3 \phi_3 \right) 
+\frac{1}{2} \lambda_\sigma (\bar{\phi}_\sigma \phi_\sigma)^2
+\frac{1}{2} \lambda_+ (\bar{\phi}_\pi \phi_\pi+\bar{\phi}_\eta \phi_\eta)^2
$$
$$
+\frac{1}{2} \lambda_- (\bar{\phi}_\pi \phi_\eta -\bar{\phi}_\eta \phi_\pi)^2
+ \lambda_2 (\bar{\phi}_\sigma \phi_\sigma)
(\bar{\phi}_\pi \phi_\pi+\bar{\phi}_\eta \phi_\eta)
$$
$$
+\lambda_3 \left[ (\bar{\phi}_\pi \phi_\pi)(\bar{\phi}_\eta \phi_\sigma)
+ (\bar{\phi}_\pi \phi_\eta)(\bar{\phi}_\pi \phi_\sigma)
+ (\bar{\phi}_\eta \phi_\pi)(\bar{\phi}_\pi \phi_\sigma)
- (\bar{\phi}_\eta \phi_\eta)(\bar{\phi}_\eta \phi_\sigma) +h.c. \right] .
\eqno(2.1)
$$
Here, for simplicity, we have denoted the case that $\phi_i$ are
SU(2)$_L$ singlets.
If the fields $\phi_i$ are SU(2)$_L$ doublets, the general form
includes, for example, $(\bar{\phi}_\sigma \phi_\pi)
(\bar{\phi}_\pi \phi_\sigma)+(\bar{\phi}_\sigma \phi_\eta)
(\bar{\phi}_\eta \phi_\sigma)$ in addition to
$(\bar{\phi}_\sigma \phi_\sigma)
(\bar{\phi}_\pi \phi_\pi+\bar{\phi}_\eta \phi_\eta)$, and so on.

The Higgs potential (2.1) cannot, in general, lead to the relation
(1.4).  Especially, the $\lambda_3$-terms badly spoil the relation
(1.4).
Only when $\lambda_3 =0$, the potential (2.1) leads to a simple
relation
$$
v_\pi^2 +v_\eta^2 = 
\frac{\lambda_2 -\lambda_\sigma}{\lambda_2-\lambda_+} v_\sigma^2 ,
\eqno(2.2)
$$
so that we get the relation
$$
v_1^2+v_2^2+v_3^2 = \frac{1}{3} K (v_1 +v_2 +v_3)^2 ,
\eqno(2.3)
$$
where
$$
K= 1+\frac{\lambda_2 -\lambda_\sigma}{\lambda_2-\lambda_+} .
\eqno(2.4)
$$
Thus, the case with
$$
\lambda_3=0, \ \ \ 
\lambda_\sigma =\lambda_+ \neq \lambda_2 ,
\eqno(2.5)
$$
can give the relation (1.4).

The Higgs potential (2.1) with the conditions (2.5) is essentially 
identical with the Higgs potential (1.5).
(Hereafter, we will use the expression (1.5) as the S$_3$-invariant
Higgs potential which can gives the relation (1.4).)
The characteristic of the Higgs potentials (1.5) [and also 
(2.1) with the constraints (2.5)] which can give the relation
(1.4) is that it is invariant under the replacement
$$
(\bar{\phi}_\sigma \phi_\sigma) \leftrightarrow 
(\bar{\phi}_\pi \phi_\pi+\bar{\phi}_\eta \phi_\eta) .
\eqno(2.6)
$$
As we stated in Sec.1, the potential (1.5) cannot give the
difference between $v_\pi$ and $v_\eta$ and it generates
a massless scalar, because the potential (1.5) is invariant
under an SU(2)-flavor symmetry for the basis $(\phi_\pi, \phi_\eta)$.
We have to introduce a symmetry breaking of the SU(2)-flavor.

\vspace{5mm}
{\large\bf 3 \ S$_3$ symmetry breaking in the Higgs potential}

In this section, we investigate an S$_3$ symmetry breaking term
which does not spoil the relation (1.4).

By the way, when we define parameters $z_i$ by $v_i =z_i v$ with the
normalization condition $z_1^2 +z_2^2 +z_3^2=1$, the parameters
$z_i$ which satisfy the relation
$$
z_1^2+z_2^2+z_3^2=1=\frac{2}{3}\left( z_1 +z_2 +z_3 \right)^2 ,
\eqno(3.1)
$$
are explicitly expressed as follows \cite{KT96}:
$$
z_1 = \frac{1-\sqrt{1-\varepsilon}}{\sqrt{6}}, \ \ 
z_2 = \frac{2+\sqrt{1-\varepsilon}-\sqrt{3}\sqrt{1+\varepsilon}}{2\sqrt{6}}, 
\ \ 
z_3 = \frac{2+\sqrt{1-\varepsilon}+\sqrt{3}\sqrt{1+\varepsilon}}{2\sqrt{6}}, 
\eqno(3.2)
$$
where the expression (3.2) has been so taken as to give $m_e \rightarrow 0$
in the limit of $\varepsilon \rightarrow 0$, and the value of
$\varepsilon$ is 
$$
\varepsilon = 0.079072 ,
\eqno(3.3)
$$
from the observed values of the charged lepton masses.
Also, for the parameters $z_a=v_a/v$ ($a=\pi, \eta, \sigma$),
we obtain
$$
z_\pi = -\frac{1}{4}\left( \sqrt{3} \sqrt{1-\varepsilon}
-\sqrt{1+\varepsilon} \right) , \ \ 
z_\eta = -\frac{1}{4}\left( \sqrt{3} \sqrt{1+\varepsilon}
+\sqrt{1-\varepsilon} \right) , \ \ 
z_\sigma = \frac{1}{\sqrt{2}} .
\eqno(3.4)
$$
We must to seek for the potential form which gives 
$z_\pi \neq z_\eta$ keeping the relation (3.1).

As an example of such an S$_3$ symmetry breaking term, in 
Ref.\cite{Koide99}, a term
$$
V_{SB} = \lambda_{SB} \left[ \xi_\pi (\bar{\phi}_\pi \phi_\pi)
- \xi_\eta (\bar{\phi}_\eta \phi_\eta) \right]^2 ,
\eqno(3.5)
$$
with $\xi_\pi \neq \xi_\eta$ has been suggested.
However, in this paper, we would like to consider a case that 
the symmetry is softly broken.
Therefore, in this section, we consider a symmetry breaking
in the mass term in the Higgs potential (1.5).

The Higgs potential is invariant under the SU(2)-flavor symmetry
for the basis $(\phi_\pi, \phi_\eta)$, i.e.
under the transformation
$$
\left(
\begin{array}{c}
\phi'_\pi \\
\phi'_\eta \\
\end{array} \right) = \left(
\begin{array}{cc}
c & -s \\
s & c \\
\end{array} \right)  \left(
\begin{array}{c}
\phi_\pi \\
\phi_\eta \\
\end{array} \right),
\eqno(3.6)
$$
where $s=\sin\theta$ and $c=\cos\theta$.
Now, we want to fix the mixing (3.6) to a special basis.
Therefore, we take a symmetry breaking form
$$
V_{SB} = \mu_{SB}^2 (\bar{\phi}'_\pi \phi'_\pi)
= \mu_{SB}^2 (c \bar{\phi}_\pi  -s \bar{\phi}_\eta )
(c {\phi}_\pi  -s {\phi}_\eta ) .
\eqno(3.7)
$$
This does not mean that the SU(2)-flavor invariant
potential (1.5) is also given in terms of the new basis
$(\phi'_\pi, \phi'_\eta, \phi_\sigma)$.
We assume that the potential $V$, Eq.~(1.5), (hereafter, 
we call it $V_0$) is still given in terms of the basis
$(\phi_\pi, \phi_\eta, \phi_\sigma)$, while, only for $V_{SB}$,
it is given by the form (3.7).
Therefore, we regard the Higgs potential
$$
V = V_0 +V_{SB} 
\eqno(3.8)
$$
as a function of the fields $(\phi_\pi, \phi_\eta, \phi_\sigma)$,
and the mixing parameter $\theta$ as a fundamental parameter 
in the present model.

Then, we obtain
$$
\left[ \mu^2 +\lambda_1 (|v_\pi|^2 +|v_\eta|^2 +|v_\sigma|^2)
+ \lambda_2 |v_\sigma|^2 \right] v_\pi 
+\mu^2_{SB} c (c v_\pi - s v_\eta) =0 ,
\eqno(3.9)
$$
$$
\left[ \mu^2 +\lambda_1 (|v_\pi|^2 +|v_\eta|^2 +|v_\sigma|^2)
+ \lambda_2 |v_\sigma|^2 \right] v_\eta 
-\mu^2_{SB} s (c v_\pi - s v_\eta) =0 ,
\eqno(3.10)
$$
$$
\left[ \mu^2 +\lambda_1 (|v_\pi|^2 +|v_\eta|^2 +|v_\sigma|^2)
+ \lambda_2 (|v_\pi|^2 +|v_\eta|^2) \right] v_\sigma = 0 ,
\eqno(3.11)
$$
from the conditions $\partial V/\partial \bar{\phi}_\pi =0$,
$\partial V/\partial \bar{\phi}_\eta =0$, and
$\partial V/\partial \bar{\phi}_\sigma =0$, respectively.
Therefore, for $\lambda_1 \neq 0$, $\lambda_2 \neq 0$, and
$\mu^2_{SB} \neq 0$, we obtain the relations
$$
|v_\pi|^2 +|v_\eta|^2  =|v_\sigma|^2 ,
\eqno(3.12)
$$
and
$$
\frac{s}{c} = \frac{v_\pi}{v_\eta} .
\eqno(3.13)
$$
The relation (3.12) leads to the relation (1.4).
The relation (3.13) means
$$
\tan\theta =\frac{\sqrt{3}(z_2-z_1)}{2 z_3 -z_2 -z_1}
=\frac{\sqrt{3}(\sqrt{m_\mu}-\sqrt{m_e})}{2 \sqrt{m_\tau} 
-\sqrt{m_\mu} -\sqrt{m_e}},
\eqno(3.14)
$$
which gives a numerical result
$$
\theta = 12.7324^\circ, \ \ {\rm i.e.} \ 
\sin\theta = 0.220398 .
\eqno(3.15)
$$
Note that the mixing parameter value (3.15) is 
in an excellent agreement with the observed Cabibbo mixing 
angle, i.e. $|V_{us}|= 0.2200\pm 0.0026$ \cite{PDG04}.
At present, this is an accidental coincidence, because
we have not yet discussed a model of the quark mixing.
(A formula for the Cabibbo angle similar to Eq.~(3.14)
is found in Ref.~\cite{KoidePRL81}.)
However, this coincidence will become a hint for seeking 
for the quark mixing model.
In the present stage, the value (3.15) of $\theta$ is
only a phenomenological result from the observed charged
lepton mass spectrum.

\vspace{5mm}
{\large\bf 4 \ Concluding remarks}

In conclusion, we have investigated a Higgs potential 
which gives the VEV structure (1.4). 
We have considered the following scenario for the
Higgs potential:

\noindent
(i) First, we consider SU(3)-flavor symmetric Higgs
potential.

\noindent
(ii) The SU(3) symmetric Higgs potential is broken by
the $\lambda_2$-terms as shown in Eq.~(1.5), but
it is still invariant under the S$_3$ flavor symmetry.

\noindent
(iii) The S$_3$ symmetric potential (1.5) is softly broken by
the term (3.7) with a phenomenological mixing parameter
$\theta$.

\noindent
Then, we can obtain a realistic charged lepton mass
spectrum for the parameter value (3.15).
We consider that those symmetry breakings are explicitly
broken at a high energy scale.
In the present low energy phenomenology, we do not discuss
the origin of those symmetry breakings.

The present model is a multi-Higgs model, so that the model
basically induces the flavor-changing neutral currents 
(FCNC).  However, the Higgs scalars $\phi_i$ in the present 
model do not couple to the quarks and leptons directly.
Symbolically speaking, in the seesaw mass matrix model
$M_f = m_L M_F^{-1} m_R$, the FCNC effects through the 
exchange of the Higgs scalars $\phi_{Li}$ are suppressed by
the order of $(M_F^{-1} m_R)^2$.
Therefore, we will be able to avoid the FCNC problem
from the present seesaw model.

Finally, we would like to emphasize that the relation (1.4) 
can be obtained independently of the explicit values of the
parameters $\lambda_1$, $\lambda_2$, and $\mu^2_{SB}$
in the Higgs potential.
The relation (1.4) is determined only by the form 
(parameter-independent structure) of the Higgs potential.
The explicit values of $z_i$ are dependent only on the 
value of $\varepsilon$ (so that on the value of the
mixing parameter $\theta$).
The magnitude of the parameter $\varepsilon$ of the S$_3$ 
violation should be understood from more fundamental theory 
in future.
We believe that the charged lepton mass spectrum will be 
described only in terms of fundamental constants without
adjustable parameters, while quark and neutrino mass
matrices will be described in terms of such fundamental
constants and some phenomenological parameters.

\vspace{5mm}
\centerline{\bf Acknowledgments}

The author would like to thank J.~Kubo and S.~Kaneko
for helpful discussion on discrete symmetries.


\newpage

\vspace{-2cm}
\hspace*{4cm}
\begin{picture}(300,120)(50,50)
\put(0,50){\thicklines \vector(1,0){40}}
\put(40,50){\thicklines \line(1,0){35}}
\put(35,30){$\bar{5}_{Li}$}
\put(75,50){\thicklines \line(0,1){5}}
\put(75,60){\thicklines \line(0,1){5}}
\put(75,70){\thicklines \line(0,1){5}}
\put(75,80){\thicklines \line(0,1){5}}
\put(70,85){\thicklines \line(1,1){10}}
\put(80,85){\thicklines \line(-1,1){10}}
\put(75,50){\circle*{5}}
\put(65,105){$\langle \phi_i \rangle$}
\put(150,50){\thicklines \vector(-1,0){40}}
\put(110,50){\thicklines \line(-1,0){35}}
\put(115,30){$5'_{Li}$}
\put(150,50){\circle*{5}}
\put(150,50){\thicklines \line(0,1){5}}
\put(150,60){\thicklines \line(0,1){5}}
\put(150,70){\thicklines \line(0,1){5}}
\put(150,80){\thicklines \line(0,1){5}}
\put(145,85){\thicklines \line(1,1){10}}
\put(155,85){\thicklines \line(-1,1){10}}
\put(150,50){\circle*{5}}
\put(140,105){$\langle 5_H \rangle$}
\put(150,50){\thicklines \vector(1,0){40}}
\put(190,50){\thicklines \line(1,0){35}}
\put(185,30){$ \overline{10}'_{Li} $}
\put(225,50){\thicklines \line(0,1){5}}
\put(225,60){\thicklines \line(0,1){5}}
\put(225,70){\thicklines \line(0,1){5}}
\put(225,80){\thicklines \line(0,1){5}}
\put(220,85){\thicklines \line(1,1){10}}
\put(230,85){\thicklines \line(-1,1){10}}
\put(225,50){\circle*{5}}
\put(215,105){$\langle \phi_i \rangle$}
\put(300,50){\thicklines \vector(-1,0){40}}
\put(260,50){\thicklines \line(-1,0){35}}
\put(265,30){$10_{Li}$}
\end{picture}

\vspace*{1cm}
\begin{quotation}
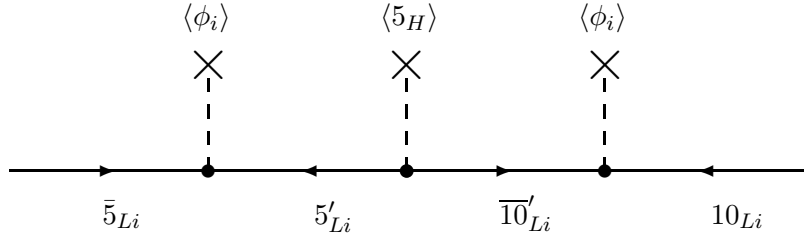
{\small\bf Fig.~1 \ Seesaw mass-generation of the charged leptons, 
where flavor-triplet scalars $\phi_i$ are singlets of SU(5).
}
\end{quotation}
\vspace{3mm}

\end{document}